\documentclass[prl,twocolumn,showpacs]{revtex4-1}
\usepackage{graphicx}
\begin{document}
\title{Increasing Quantum Degeneracy by Heating a Superfluid}
\author{D.J. Papoular$^{1}$, G. Ferrari$^{1}$, L.P. Pitaevskii$^{1,2}$,
  S. Stringari$^{1}$}
\affiliation{${}^{1}$INO-CNR BEC Center and Dipartimento di Fisica, 
  Universit\`a di Trento, 38123 Povo, Italy}
\affiliation{${}^{2}$Kapitza Institute for Physical Problems, Kosygina
2, 119334 Moscow, Russia}
\begin{abstract}
We consider a uniform superfluid confined in two compartments connected by a superleak and initially held at equal temperatures. If one of the two compartments is heated, a fraction of the superfluid will flow through the superleak. We show that, under certain thermodynamic conditions, the atoms flow from the hotter  to the colder compartment, contrary to what happens in the fountain effect observed in superfluid Helium.
This flow causes quantum degeneracy to increase in the colder compartment. 
In superfluid Helium, this novel thermomechanical effect takes place in the phonon regime of very low  temperatures. In dilute quantum gases, it
occurs at all temperatures below $T_c$.
The increase in quantum degeneracy reachable through the adiabatic displacement of the wall separating the two compartments  is also discussed.
\end{abstract}


\maketitle

The thermomechanical effect is an important manifestation of superfluidity. 
It has historically been observed via the fountain effect, i.e. the increase in the pressure in a narrow tube, one of whose ends dips in a 
bath of superfluid Helium 4, when the tube is heated
\cite{allen_b:Nature1938}.
It has recently also attracted interest in the context of dilute ultracold atomic gases
\cite{marques:PRA2004,karpiuk:arxiv2010} as a potential signature of superfluidity in these systems. These works have focused on the dynamical aspects of the phenomenon, using the hydrodynamic \cite{marques:PRA2004} or the classical field \cite{karpiuk:arxiv2010} approach. 

The purpose of the present work is to investigate the thermomechanical effect
by exploiting the conditions imposed by equilibrium thermodynamics, pointing 
out novel  features exhibited by superfluids in properly chosen thermodynamic 
regimes.
The experiment we propose is reminiscent of the original fountain effect 
\cite{allen_b:Nature1938}, with one important difference.
In the original experiment, the pressure at the surface of the liquid Helium 
bath is constantly equal to the saturated vapor pressure, and the 
height growth of the liquid in the narrow tube is determined 
by the equilibrium between gravity
and the pressure increase $\delta p=s\delta T$ of the liquid near the 
superleak. Here $\delta T$ is the temperature difference between the bath and 
the tube, and $s$ is the entropy per unit volume in the liquid phase.
In the situation considered in this paper, the quantum fluid instead 
occupies a fixed total volume, and the flow of particles through the superleak 
is caused by the density difference produced by the heating process. 
Consequently, it is determined by the compressibility of the fluid.
In the case of liquid Helium, the compressibility is small;
it is much larger in dilute quantum gases. We predict that, in the phonon 
regime of  superfluid Helium, and for all temperatures below $T_c$ in the case 
of dilute gases, atoms flow through the superleak from the hotter to the 
colder region,
contrary to what happens in the fountain effect, and
resulting in an increase of quantum degeneracy in the colder compartment. 
The cooling mechanism proposed in the present paper, based on a filtering process 
through the superleak, differs from other adiabatic cooling mechanisms 
considered in ultracold atomic gases, like the adiabatic formation of 
Bose-Einstein condensation with non--harmonic traps \cite{stamperkurn:PRL1998} 
or the entropy exchange in mixtures of different atomic species 
\cite{catani:PRL2009}.

\begin{figure} 
    \includegraphics[width=.4\linewidth]
    {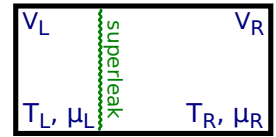}
    \caption{\label{fig:expschematics}
    Schematics of the proposed experiment. The left and right compartments, initially in equilibrium, have constant volumes and are filled with a homogeneous quantum fluid. Heating the right compartment, or displacing the superleak, causes the system to evolve towards a new thermodynamic state satisfying chemical, but not thermal, equilibrium.}
\end{figure}

We consider two compartments, hereafter called left ($L$) and right ($R$) compartments,  
filled with a homogeneous superfluid (liquid Helium 4 or a dilute atomic gas) and connected via a superleak, which only allows for the transmission of the superfluid component (see Fig.~\ref{fig:expschematics}).
Initially, the superfluids occupying the two compartments have the same temperature ($T_L^0=T_R^0=T^0$)  and the same chemical potential ($\mu_L^0=\mu_R^0$). 
If the right compartment is heated, the system will eventually reach a new equilibrium configuration characterized by equal chemical potentials, but different temperatures. The equilibrium between the final--state chemical potentials is ensured by the flux of the superfluid through the superleak.
However, the condition of equal temperatures cannot be ensured because the superfluid component does not carry any entropy.

By calling $\delta T =T_R-T_L$,  $\delta n=n_R-n_L$ and $\delta p=p_R-p_L$ the small differences between the final temperatures, densities and pressures of the two  compartments, and imposing equal chemical potentials 
$\delta \mu=\mu(T_R,n_R)-\mu(T_L,n_L)=0$, we can write:
\begin{equation} \label{eq:lindeviations_deltan}
  \delta n=
  \left.\frac{\partial n}{\partial T}\right|_\mu \delta T
  \quad \text{ and } \quad
  \delta p=
  \left.\frac{\partial p}{\partial T}\right|_\mu \delta T
  \ ,
\end{equation}
both partial derivatives being evaluated at the initial equilibrium parameters. 
The Gibbs--Duhem relation 
for a homogeneous system implies ${\partial p}/{\partial T}|_\mu=s$,
where $s=S/V>0$ is the entropy per unit volume. As a consequence the pressure $p_R$ in the warmer compartment is always higher than $p_L$. The density difference $\delta n$ can instead exhibit a different behavior.
In fact, one can write:
\begin{equation} \label{eq:deltan_general}
  \left.\frac{\partial n}{\partial T}\right|_\mu = -\left.\frac{\partial n}{\partial \mu}\right|_T\left.\frac{\partial \mu}{\partial T}\right|_n=
  - n^2\, \kappa_T \,
  \left.\frac{\partial \mu}{\partial T}\right|_n
  \quad ,
\end{equation}
the  isothermal  compressibility $\kappa_T$ being a positive  quantity. 
Hence, the direction of the flow is dictated by the sign of
${\partial\mu}/{\partial T}|_n$. If this derivative is negative,
and thus $\partial n/\partial T|_\mu>0$,
the atoms flow from the colder to the warmer compartment (positive flow).
The opposite happens if
${\partial\mu}/{\partial T}|_n>0$, and hence $\partial n/\partial T|_\mu<0$
(negative flow).
In the latter case, the quantum degeneracy of the colder compartment 
increases. In fact,
the total entropy in the left compartment remains constant, but the entropy per particle $S_L/N_L$ decreases. 
In order to ensure adiabaticity, the heating process (as well as the displacement of the separating wall discussed in the last section) should be slow enough to ensure thermalization in the left compartment, and the velocity $v$ of the atoms going through the superleak during the process should be smaller than a critical value dictated by the geometry of the experiment. This condition will also ensure the absence of dissipation due to the creation of vortices. If the transmission of the atoms relies on quantum tunneling, the current $nv$ should be smaller than the critical Josephson current. The actual implementation of these conditions will be discussed elsewhere.

Using the Gibbs--Duhem relation $dp=sdT+nd\mu$,
and introducing the thermal expansion coefficient 
$n\alpha_p=-{\partial n}/{\partial T}|_p$,
Eq.~(\ref{eq:deltan_general}) can also be written in the form
\begin{equation}\label{eq:n_partialderiv_T}
  \left.\frac{\partial n}{\partial T}\right|_\mu =
  T(\kappa_T \,s - \alpha_p) \, \frac{n}{T}
  \ .
\end{equation}
This form is particularly useful in the case of superfluid Helium where the quantities
$\kappa_T$, $s$ and $\alpha_p$ are experimentally available.  
It explicitly shows that the direction of the flow depends on the sign of $(\kappa_T s-\alpha_p)$. 

A general result concerning the behavior of the thermodynamic quantity 
$\left.\partial \mu /\partial T\right|_n$ entering Eq.~(\ref{eq:deltan_general}) can be inferred in the very low temperature regime, where the thermodynamic behavior of a superfluid is governed by the thermal excitation of phonons (phonon regime). In this regime, the chemical potential behaves as $\mu(n,T)=\mu_0(n)+\mu_\mathrm{phon}(n,T)$ with 
\begin{equation}\label{eq:mu_phonons}
\mu_\mathrm{phon}(n,T) =
          \frac{\pi^2}{30}\frac{(k_BT)^4}{\hbar^3 c_0^4}
          \left.\frac{\partial c_0}{\partial n}\right|_T
          \ ,
\end{equation}
while $\mu_0(n)$ and $c_0(n)$ are the $T=0$ values of the
chemical potential and the sound velocity, respectively. In all known superfluids the quantity $\partial c_0 / \partial n|_T$ is positive. 
Therefore, 
in the phonon regime one always has $\partial \mu/\partial T|_n>0$,
and hence a negative flow.

We now discuss in a more systematic way the temperature dependence of 
$ \left.\partial n /\partial T\right|_\mu$
in the case of superfluid Helium and  of dilute quantum gases.

\begin{figure*}
  \begin{minipage}{.325\linewidth}
    \includegraphics[angle=-90,width=\linewidth]
    {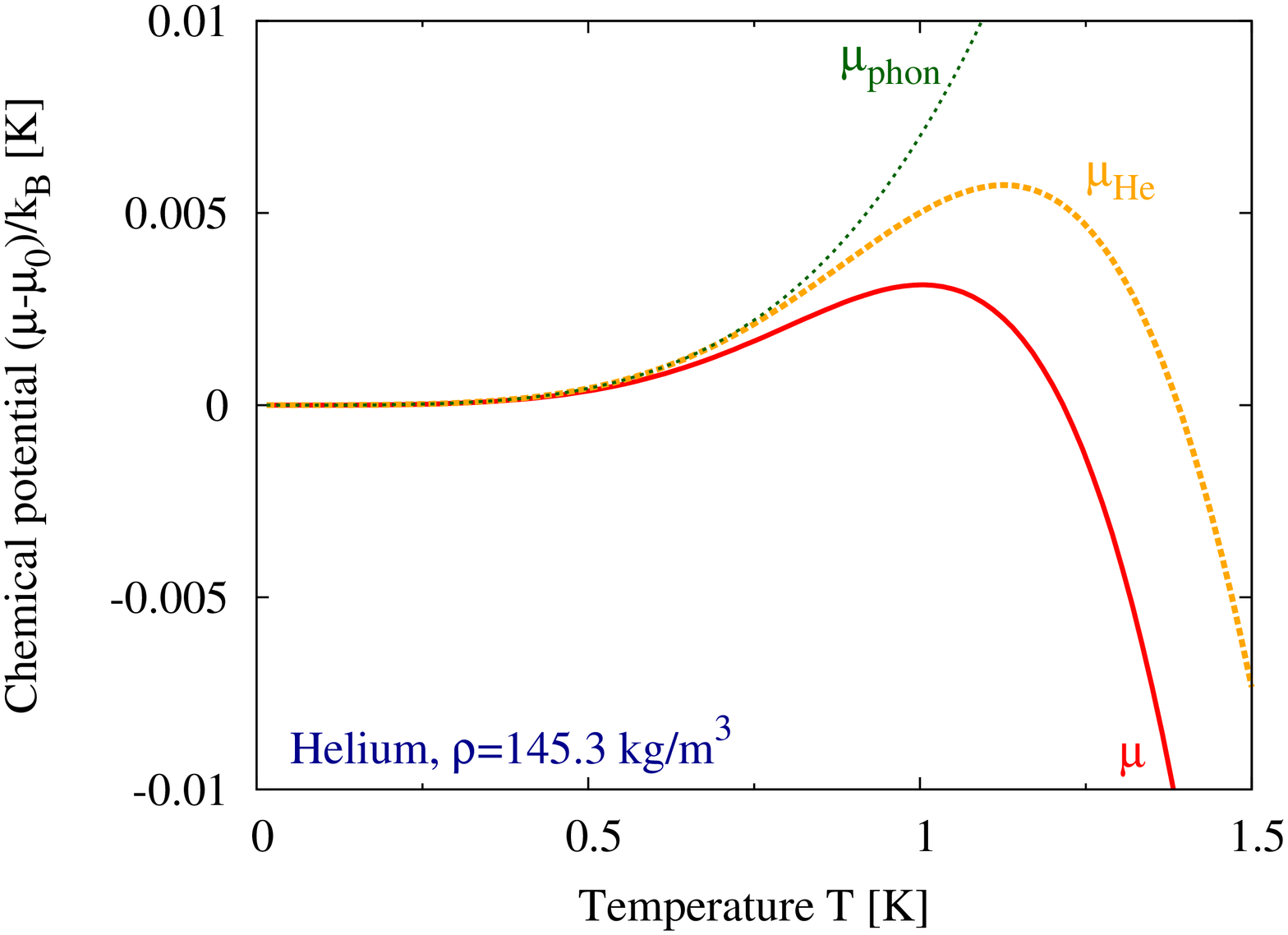}
  \end{minipage}
  \begin{minipage}{.325\linewidth}
    \includegraphics[angle=-90,width=\linewidth]
    {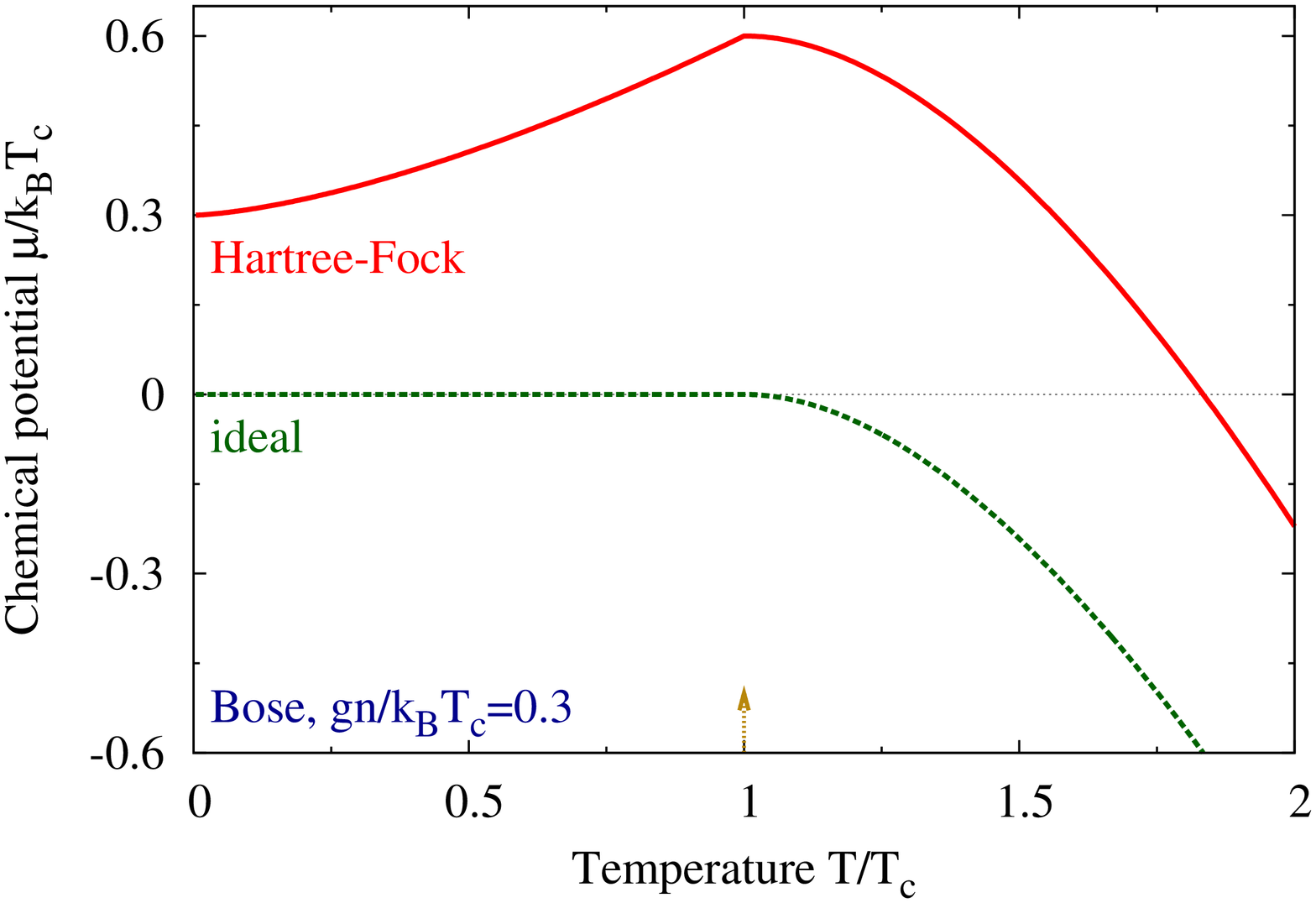}
  \end{minipage}
  \begin{minipage}{.325\linewidth}
    \includegraphics[angle=-90,width=\linewidth]
    {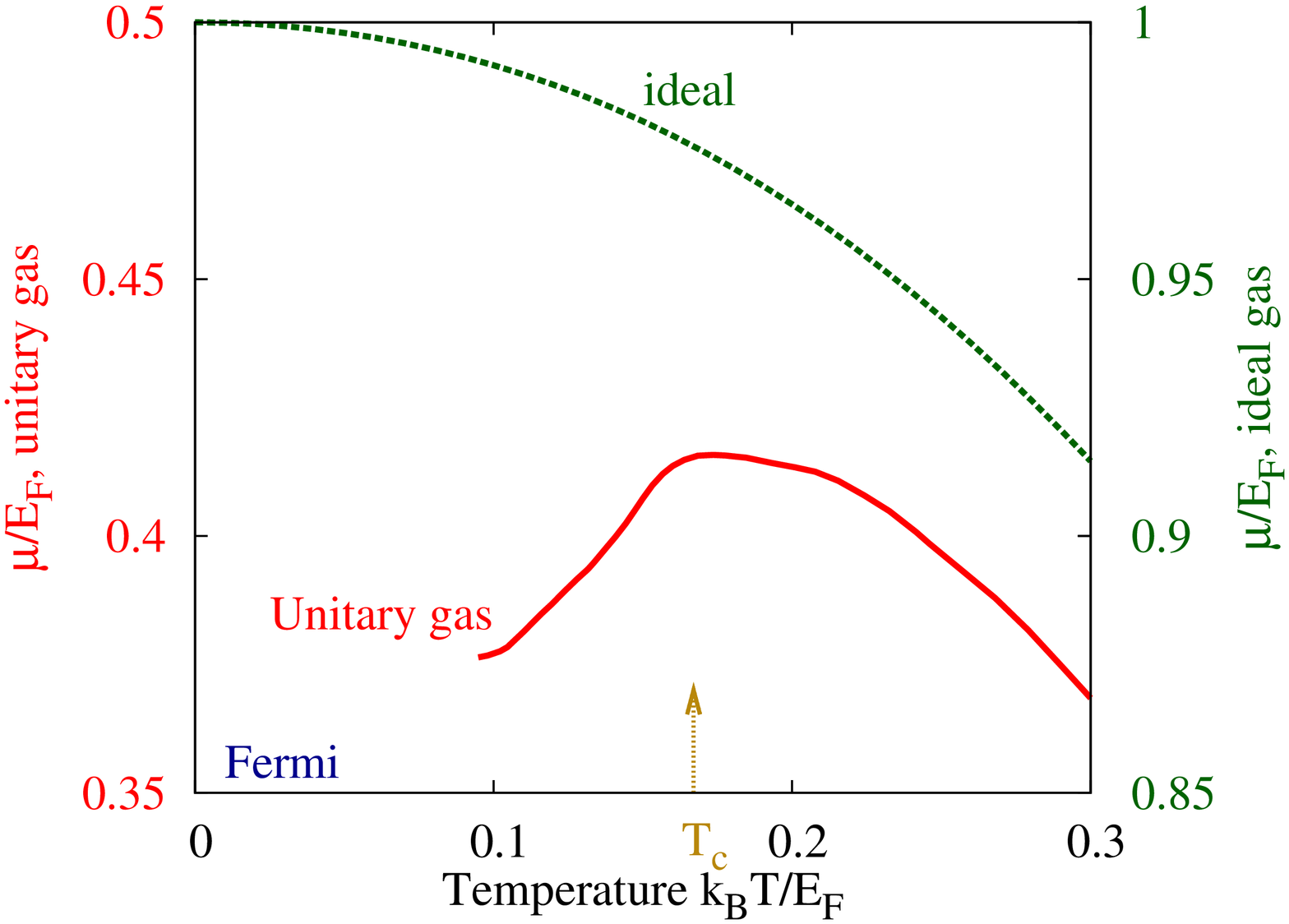}
  \end{minipage}
  \caption{ \label{fig:muIncreaseT}
    Chemical potential as a function of T for three different superfluids. 
    \textbf{Left:} superfluid ${}^4\mathrm{He}$,
    calculated using the equation of state in \cite{arp:IJT2005} (red).
    The orange curve is the sum of the phonon and roton contributions
    (Eqs.~(\ref{eq:mu_phonons}) and (\ref{eq:mu_phonons_rotons})); 
    the phonon contribution is also shown separately (green).
    \textbf{Middle}: ultracold Bose gas (red), calculated from the Hartee--Fock
    prediction (Eq.~(\ref{eq:EoS_HF}))
    with $gn/k_BT_c=0.3$. 
    \textbf{Right:} ultracold Fermi gas at the unitary limit (red)
    \cite{ku:arXiv2011}.
    For both gaseous superfluids, the ideal gas prediction
    is also shown (green).
  }
\end{figure*}

\textbf{Superfluid Helium}. When one increases the temperature and leaves the phonon regime, the thermodynamic behavior of superfluid helium is soon dominated by the thermal excitation of rotons which deeply affect the behavior of the chemical potential. In Fig.~\ref{fig:muIncreaseT} (left) we show the temperature dependence of $\mu(T)$ extracted from the measured thermodynamic functions of Helium \cite{arp:IJT2005} for temperatures below $1.5\,\mathrm{K}$.
Its variation from the $T=0$ value at zero pressure,
$\mu_0/k_B=-7.16\,\mathrm{K}$, is small.
However, its change of behavior when $T$ increases,
and in particular the change of sign of 
$ \left.\partial \mu /\partial T\right|_n$, is apparent. This effect is 
caused by the roton contribution to thermodynamics which can be written in the 
form   
\begin{equation}\label{eq:mu_phonons_rotons}
\mu_\mathrm{rot}(n,T)=
          f\,
          (k_BT)^{1/2} \,
          \left.\frac{\partial\Delta}{\partial n}\right|_T \,
          e^{-\Delta/k_B T}
          \ .
\end{equation}
In Eq.~(\ref{eq:mu_phonons_rotons}) we have defined
$f=2p_0^2\sqrt{m^*}/(2\pi\hbar^2)^{3/2}$, where
$p_0$ is the momentum corresponding to the roton minimum in the excitation spectrum and $m^*$ is the corresponding effective mass \cite{wilks:Clarendon_1967}.
The coefficient $f$ depends only weakly on $n$ \cite{arp:IJT2005}
and,  for simplicity, in  Eq.~(\ref{eq:mu_phonons_rotons})  we have only retained the density dependence of 
the roton gap $\Delta(n)$ which is well known to decrease with increasing $n$, thereby providing a negative contribution  to 
$ \left.\partial \mu /\partial T\right|_n$.
The approximate expression
$(\mu_\mathrm{phon}+\mu_\mathrm{rot})$ correctly reproduces the qualitative
behaviour of the full thermodynamic result (see Fig.~\ref{fig:muIncreaseT} left), which
confirms the key physical role played by rotons.
The curves refer to the value  
$\rho=m_\mathrm{He} n=145.3\,\mathrm{kg\, m^{-3}}$ for the density,  $m_\mathrm{He}$ being the the mass of a single Helium atom. 
The analysis reveals that, for temperatures larger than $\sim 1\,\mathrm{K}$,  superfluid Helium is charaterized by
$\partial n/\partial T|_\mu>0$, i.e. by a positive flow through the superleak.
In this regime, the thermal expansion coefficient is actually negative 
\footnote{For liquid Helium 4, at saturated vapour pressure \cite{donnelly:JPCRD1998}, $\alpha_p<0$ for
$1\,\mathrm{K}<T<2.2\,\mathrm{K}$.} 
and the two terms in Eq.~(\ref{eq:n_partialderiv_T})
add up with the same sign.
Using the values of the thermodynamical functions given in 
\cite{arp:NIST1998,brooks:JPCRD1977}, we predict that the relative amplitude
$(T/\rho)\partial\rho/\partial T|_\mu$ is of the order of $10^{-2}$ for $T=1.8\,\mathrm{K}$.

\begin{figure*}
  \begin{minipage}{.45\textwidth} \includegraphics[angle=-90,width=.7\textwidth] {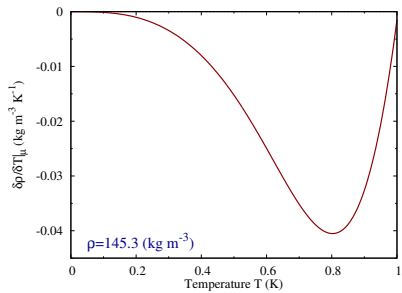}
  \end{minipage}
  \begin{minipage}{.45\textwidth}
    \caption{\label{fig:helium_phononregime_UnitT_2D}
      Amplitude 
      $\left.\frac{\partial\rho}{\partial T}\right|_\mu$
      of the thermomechanical effect in liquid ${}^4\mathrm{He}$:
      dependence on the temperature $T$ for the fixed 
      density $\rho=145.3\,\mathrm{kg/m^3}$.
    }
  \end{minipage}
\end{figure*}

For smaller temperatures, when we approach the phonon regime,
Helium exhibits a negative flow throw the superleak.
The temperature where 
this effect is largest depends on the density,
and decreases as the density increases.
For the density $\rho=145.3\,\mathrm{kg/m^3}$,
close to the lowest density for which Helium 4 is liquid,
the relative amplitude  of the predicted effect is largest for $T=0.8\,\mathrm{K}$  (see Fig.~\ref{fig:helium_phononregime_UnitT_2D})
where we find 
$(T/\rho)\partial\rho/\partial T|_\mu \approx -2\cdot 10^{-4}$. This value is  
two orders of magnitude smaller than the  typical  relative amplitude corresponding to  thermomechanical effect in the roton region.
The reason for this large difference 
can easily be understood from
Eq.~(\ref{eq:n_partialderiv_T}):
in the phonon regime, $\kappa_T s$ and $-\alpha_p$ have opposite signs,
leading to a strong suppression of  the total amplitude. Furthermore, both $s$ and $\alpha_p$ become smaller and smaller  as $T\to 0$. 
Although small, the novel thermomechanical effect in the phonon regime of Helium should be large enough to be detected experimentally by measuring the change in the index of refraction 
or the dielectric constant $\epsilon$
of the liquid in the left compartment. Indeed, it should lead to a relative variation $\delta\epsilon/\epsilon \approx 10^{-6}$
\cite{donnelly:JPCRD1998}, which is well above the typical experimental sensitivity \cite{chan:JLTP_1977}. It is worth pointing out that the thermomechanical effect discussed above is based on the assumption that the total volume $V_L+V_R$ occupied by the fluid is kept constant. This effect should not be confused with the fountain effect, which occurs in situations where there is no constraint on the total volume. This latter effect
is driven by the pressure difference
$s\delta T$ and, hence, is always characterized by a positive flow through the superleak.

\textbf{Dilute quantum gases.}
The thermodynamic behavior of quantum gases is very different from the one of superfluid Helium, due to the absence of the roton minimum in the excitation spectrum. As soon as one leaves the phonon regime described by Eq.~(\ref{eq:mu_phonons}), the relevant excitations are of single--particle nature. In the case of dilute  Bose--Einstein condensates, a reliable  description of the single--particle regime is provided  by Hartree--Fock theory, which yields the following expression for the chemical potential in the superfluid regime below $T_c$ \cite{pitaevskii:BEC2003}: 
\begin{equation} \label{eq:EoS_HF}
  \mu(n,T)=g\left(n_0 + 2n_T\right)=g\left[
    n
    +\zeta(3/2) 
    \left( \frac{m k_B T}{2\pi\hbar^2}\right)^{3/2}
    \right]
    \ ,
\end{equation}
and the expression $\mu=\mu_\mathrm{id}+2gn$  for $T>T_c$,
$\mu_\mathrm{id}$ being the chemical potential of the ideal gas.
In Eq.~(\ref{eq:EoS_HF}),
$n_0$  and $n_T$ are the densities of the condensate and of the thermal components respectively,
$g=4\pi\hbar^2 a/m$ is the coupling constant ($a$ is  the scattering length characterizing the atom--atom interactions  and  $m$  the mass of a single atom),
and $\zeta(3/2)\approx 2.612$.
Consistently with the  weak coupling scheme, 
in the second equality of Eq.~(\ref{eq:EoS_HF})
we have used the ideal gas expression 
$n_T=\zeta(3/2) (m k_B T/2\pi\hbar^2)^{3/2}$ 
for the thermal density
\footnote{The number of thermal atoms depending on $T$ but not on $N$ is characteristic of the ideal gas. The violation of this saturation property has recently been investigated experimentally \cite{tammuz:PRL_2011}}.
Equation (\ref{eq:EoS_HF}) shows that the chemical potential is an increasing function of temperature up to the critical temperature 
$mk_B T_c=2\pi\hbar^2 (n/\zeta(3/2))^{2/3}$, 
where $\mu$ reaches its maximum which is twice
the $T=0$ value  \footnote{The applicability of Hartree-Fock theory actually breaks down when one approaches $T_C$}. 
The temperature dependence of the chemical potential at fixed density 
can now be 
measured in quantum gases, where the equation of state is obtained by suitable 
integration of the density profiles of the trapped gas  
or direct {\it in situ} measurements.  
These measurements have so far been carried out for Fermi gases (see Fig.~\ref{fig:muIncreaseT} (right)) \cite{ku:arXiv2011} but can also be performed in Bose gases \cite{salomon:2011}.

Combining Eqs.~(\ref{eq:deltan_general}) and (\ref{eq:EoS_HF}), we obtain the following expression for the relative  amplitude of the thermomechanical effect in the Bose  gas below $T_c$:
\begin{equation} \label{eq:deltan_gas_HF}
  \frac{T}{n} \left.\frac{\partial n}{\partial T}\right|_\mu = 
  -\frac{3}{2} \, 
  \left( \frac{T}{T_c} \right)^{3/2} \,
  \ ,
\end{equation}
yielding the negative flow of particles through the superleak
for all temperatures\footnote{Note that our prediction~(\ref{eq:deltan_gas_HF}) for the negative flow holds for a homogeneous Bose gas. In the presence of a harmonic trapping potential, the chemical potential is a decreasing function of $T$ \cite{pitaevskii:BEC2003}, and Eq.~(\ref{eq:deltan_general}) shows that, in this case, a positive flow through the superleak will be observed, in accordance with the numerical simulation  reported in \cite{karpiuk:arxiv2010}.}. 
Equation (\ref{eq:deltan_gas_HF}) also shows that the  relative amplitude  is of order $1$ for temperatures of the order of  the critical temperature. 
The flow through the superleak being negative in the whole range of temperatures below $T_c$, and the relative amplitude of the effect being very large, are the two main differences exhibited by dilute BEC gases with respect to superfluid Helium.
In particular, the second feature is the consequence of the large compressibility of the gas. 
 
It is interesting to discuss now the thermomechanical effect in terms of the increase of quantum degeneracy in the left compartment.
Although our expression (\ref{eq:EoS_HF}) for the chemical potential accounts for the weak interaction between the atoms in the condensate,
the entropy of the gas in each compartment is well described by that of an ideal gas.
In this approximation, the entropy of a Bose--condensed gas depends on the temperature and the volume but not on the number of particles. 
Therefore, the flux of particles through the superleak, caused by heating the right compartment, changes neither the entropy $S_L$ nor the temperature $T_L$ of the left compartment.
However, as $S_L$ is distributed among a greater number of particles,
the quantum degeneracy of the left compartment increases.
Assuming that $V_R\gg V_L$, we find
$\delta (S_L/N_L)/(S_L/N_L)=-3/2(T^0/T_c^0)^{3/2}\delta T_R/T^0$, where
$T_c^0$ is the critical temperature for the initial density $n_0$.

We now discuss an even more efficient way to increase quantum degeneracy by exploiting the novel thermomechanical effect described above.
We consider the adiabatic displacement of
the wall separating the two compartments of Fig.~\ref{fig:expschematics}
towards the right. 
This displacement has two important consequences.
First,
due to adiabaticity, the temperature decreases in the left compartment and 
increases in the right compartment, the entropy of the gas being proportional 
to $VT^{3/2}$ if the effect of interactions can be neglected.
Second, 
the number of atoms in the left compartment increases, due to the presence of 
the superleak.
Note that the temperature variation in each compartment is independent of the presence of the 
superleak as the entropy does not depend on the number of particles. 
In the absence of the superleak, 
$n_L$ would decrease, and the quantum degeneracy
in both compartments would remain unchanged.
In the presence of the superleak, the situation is different:
we find that the density in the left compartment actually increases with respect to the initial value (in contrast to the decrease one would naively 
expect due to the increase of $V_L$). This follows from the thermal density being higher in the right compartment and the chemical equilibrium condition
$\mu_L=\mu_R$.
These two combined effects cause a strong increase of quantum degeneracy. 
We express it in terms of the reduced temperature
$\tau_L=T_L/T_{cL}$ in the left compartment, where the critical temperature $T_{cl}$ corresponds to the final value of $n_L$.
Assuming that $V_R$ always remains much larger than $V_L$,
the thermodynamic properties of the right compartment are not affected by the displacement of the superleak, and the final value of $\tau_L$
is given by:
\begin{equation}
  \frac{\tau_L}{\tau_0}=
  \left[
    \frac{V_L}{V_L^0}(1+\tau_0^{3/2})-\tau_0^{3/2}
  \right]^{-2/3}
  \ ,
\end{equation}
where $\tau_0$ is the initial reduced temperature,
and $V_L^0$ and $V_L$ are the initial and final 
volumes of the left compartment, respectively.
Hence, the quantum degeneracy can in principle be made arbitrarily
large by taking $V_R\gg V_L\gg V_L^0$.

The novel thermomechanical effects we have predicted for the Bose gas are also expected to occur in a Fermi gas in the superfluid regime.
The most interesting case is the unitary Fermi regime, corresponding to an infinite value of the scattering length. This system has been achieved experimentally \cite{ku:arXiv2011} and its theoretical properties are  reasonably well understood \cite{haussmann:PRA2007}. An interesting feature of this system is that its thermodynamic functions exhibit a universal behavior, independent of the value of the interaction strength.  In Fig.~\ref{fig:muIncreaseT} (right) we report the recent experimental
measurement of the chemical potential as a a function of $T/T_F$, where $k_BT_F=(\hbar^2/2m)(3\pi^2 n)^{2/3}$ is the Fermi energy defined in uniform matter, together with the prediction for the ideal Fermi gas.
Like in the case of Bose gases, the chemical potential of the Fermi superfluid is an increasing function of $T$ up to the critical value. 
Its measured behavior is reasonably well reproduced by 
the theoretical prediction obtained using the variational many--body formalism \cite{haussmann:PRA2007}. In the case of the ideal gas, the chemical potential is instead a decreasing function of $T$, revealing that the
positive slope of $\mu(T)$ is a clear  consequence of superfluidity.  
Compared to the Bose case, the relative amplitude 
$(T/n) {\partial n}/{\partial T}|_\mu$
of the thermomechanical effect is smaller in Fermi gases because of the much smaller compressibility of these systems. 

Finally, we briefly discuss how the thermomechanical effect  could be implemented in a trapped quantum gas. 
Well established tools, such as magnetic traps and optical dipole potentials, can readily be used to create the two--compartment potential \cite{andrews:Science1997}, including the implementation of box--like
traps \cite{meyrath:PRA_2005}.
The superleak can be implemented using quasi--1D optical dipole potentials with a reduced number of transverse excitation modes 
\cite{serwane:Science_2011}.
Other important issues concern the consequences of the thermomechanical effect in the presence of harmonic confinement, and in particular the optimization of the cooling mechanism through the adiabatic displacement of the wall separating the two compartments.

We are grateful to S.\ Balibar for fruitful discussions.
This work has been supported by ERC through the QGBE grant.

%

\end{document}